\documentclass[fleqn,twoside]{article}
\usepackage{espcrc2}

\usepackage{graphicx,amssymb,subfigure,psfrag}
\usepackage[figuresright]{rotating}

\newcommand{\AmS}{{\protect\the\textfont2
  A\kern-.1667em\lower.5ex\hbox{M}\kern-.125emS}}

\title{
{\vspace{-1.2em} \parbox{\hsize}{\hbox to \hsize
{\hss  \normalsize HU-EP-02/38, TRINLAT-02/04}}} \\
SO(3) Yang-Mills theory on the lattice\thanks{Talk given by A. Barresi at
Lattice2002, Boston.}}

\author{Andrea Barresi\address[HU]{Humboldt-Universit\"at zu Berlin, Institut f\"ur Physik, 10115 Berlin, Germany}, Giuseppe Burgio\address{School of Mathematics, Trinity College, Dublin 2, Ireland}, Michael M\"uller-Preussker\addressmark[HU]}

\begin{document}

\begin{abstract}
We numerically investigate the phase structure of pure $SO(3)$ LGT at zero and non-zero 
temperature in the presence of a $Z_2$ blind monopole chemical potential. The physical
meaning of the different phases, a possible symmetry breaking mechanism as well as the
existence of an order parameter for the finite temperature phase transition are discussed.
\vspace{1pc}
\end{abstract}

\maketitle

\section{INTRODUCTION}

The deconfinement phase transition, as seen on the lattice, is usually associated 
with the breaking of the global $\mathbb{Z}_N$ center symmetry in pure $SU(N)$ 
gauge theories \cite{PoSu}. Expecting universality the occurence of the transition should be 
independent
of the group representation chosen for the lattice action. A finite temperature
investigation with an $SO(3)$ Wilson action 
might offer interesting insight to the present understanding of confinement.

An $SU(2)$ mixed fundamental-adjoint action was originally studied by  Bhanot and Creutz \cite{1BC81}:
\begin{equation}
\label{eq1}
S\!=\!\sum_{P}\Bigg[\!\beta_{A}\Bigg(1-\frac{\mathrm{Tr}_{A}U_{P}}{3}\Bigg)+\beta_{F}\Bigg(1-\frac{\mathrm{Tr}_{F}U_{P}}{2}\Bigg)\!\Bigg]\,.
\end{equation}
They found the well-known non-trivial phase diagram characterized by
 first order bulk phase transition lines. A similar phase diagram is shared by $SU(N)$ theories with $N\ge 3$ \cite{2BC81}. 

Halliday and Schwimmer \cite{1HS81} found a similar phase diagram using a Villain discretization for the center blind part of action (\ref{eq1})
\begin{equation}
S=\!\!\sum_{P}\!\!\Bigg[\!\beta_{V}\Bigg(\!1-\frac{\sigma_{P}\mathrm{Tr}_{F}U_{P}}{2}\!\Bigg)\!+\beta_{F}\Bigg(\!1-\frac{\mathrm{Tr}_{F}U_{P}}{2}\!\Bigg)\!\Bigg]
\end{equation}
$\sigma_{P}$ being an auxiliary $\mathbb{Z}_2$ plaquette variable. By defining $\mathbb{Z}_2$ magnetic monopole and electric vortex densities 
$M=1-\langle\frac{1}{N_{c}}\sum_{c}\sigma_{c}\rangle$,
$E=1-\langle\frac{1}{N_{l}}\sum_{l}\sigma_{l}\rangle$ with
$\sigma_{c}=\prod_{P\epsilon\partial c}\sigma_{P}$ and
$\sigma_{l}=\prod_{P\epsilon\hat{\partial} l}\sigma_{P}$
they argued that the bulk phase transitions were caused by condensation of these lattice artifacts. They also suggested \cite{2HS81} a possible suppression mechanism via the introduction of chemical potentials of the form
$\lambda\sum_{c}(1-\sigma_{c})$ and $\gamma\sum_{l}(1-\sigma_{l})$.

For $\lambda\ge 1$ and $\gamma\ge 5$ Gavai and Datta \cite{1G99} found 
lines of second order finite temperature phase transitions crossing the 
$\beta_V$ and $\beta_F$ axes.
In the limiting case $\beta_{F}=0$ and $\gamma=0$, i.e. an $SO(3)$ theory with a 
$\mathbb{Z}_2$ monopole chemical potential, a quantitative study is difficult because 
the $\mathbb{Z}_2$ global symmetry remains unbroken and there is no obvious order parameter.
A thermodynamical approach \cite{2G99} shows a steep rise in the energy density with $N_{\tau}=2,4$ 
and a peak in the specific heat at least for $N_{\tau}=2$, supporting the idea of a second 
order deconfinement phase transition also in this case. The authors have seen the adjoint Polyakov loop $L_A$
to fluctuate around zero below the phase transition and to take the values $1$ and $-\frac{1}{3}$ 
above the phase transition as $\beta_V\to \infty$.

Jahn and de Forcrand investigated the Villain action with $\lambda=0$ at the bulk phase transition but effectively at very large T and, on the basis of previous works of Kovacs and Tomboulis \cite{KoTo} and
Alexandru and Haymaker \cite{AlHay}, suggested that the 
negative state $-\frac{1}{3}$ could be associated to a non-trivial twist sector \cite{JaDF}.   

\section{ADJOINT ACTION WITH CHEMICAL POTENTIAL}

In this paper we continue an investigation \cite{BBM} with an adjoint 
representation Wilson action modified by a chemical potential suppressing the 
$\mathbb{Z}_2$ magnetic monopoles
\begin{equation}
S=\frac{4}{3}\beta_{A}\sum_{P}\Bigg(1-\frac{\mathrm{Tr}_{F}^{2}U_{P}}{4}\Bigg)+\lambda\sum_{c}(1-\sigma_{c})\,.
\end{equation}
The link variables are taken in the fundamental representation to speed up our simulations. 
A standard Metropolis algorithm is used to update the links.
We also defined a twist observable, $k_x\equiv\frac{1}{2}\big(1-\frac{1}{L_yL_z}
\sum_{L_yL_z}N_{xt}\big)$. Both $N_{xt}\equiv\prod_{P\;\in\; \mathrm{plane}\; xt}
\mathrm{sign}(\mathrm{Tr}_F U_P)$ and
$\sigma_{c}=\prod_{P\epsilon\partial c}\mathrm{sign}(\mathrm{Tr}_{F}U_{P})$ are 
center blind, $U_{\mu}(x)\!\rightarrow\! -U_{\mu}(x)\Rightarrow
\sigma_{c}\!\rightarrow\!\sigma_{c},N_{xt}\!\rightarrow\! N_{xt}\; ,
\;\forall \mu,x,c$.  

\begin{figure}[htb]
\subfigure[twist]{
\includegraphics[width=2.75cm,angle=-90]{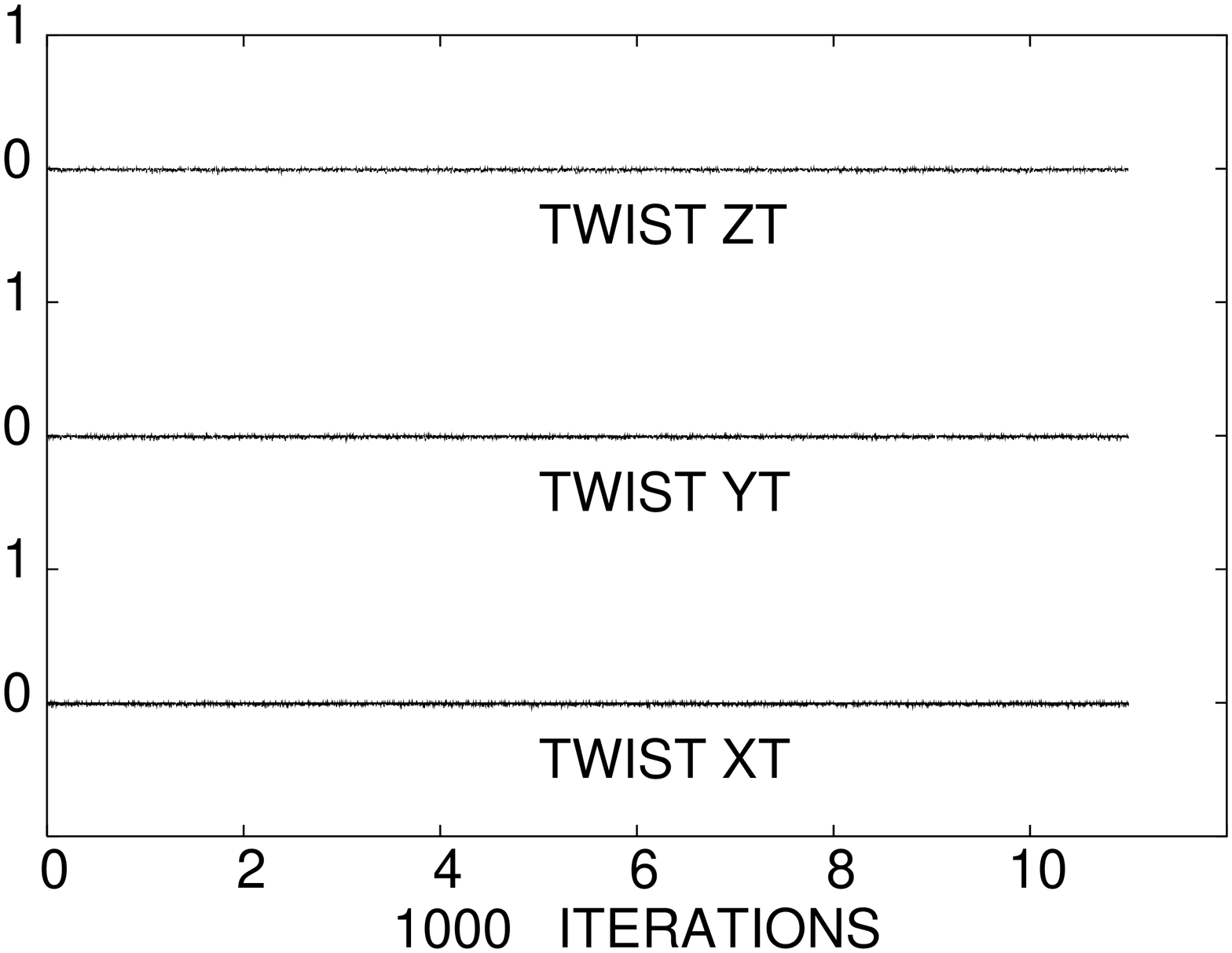}}
\subfigure[$\beta_A=0.9$]{
\includegraphics[width=2.75cm,angle=-90]{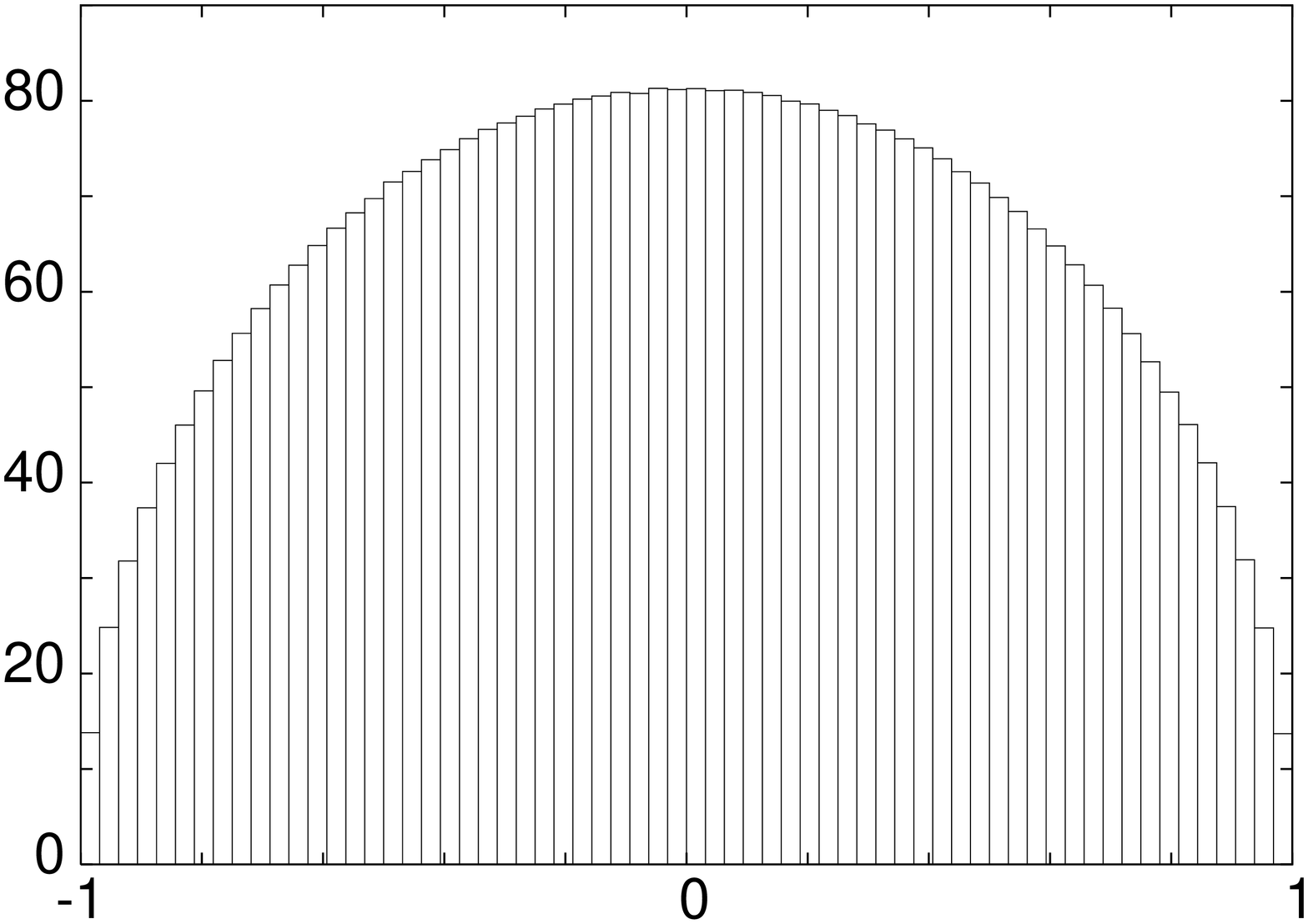}}
\subfigure[$\beta_A=1.2$]{
\includegraphics[width=2.75cm,angle=-90]{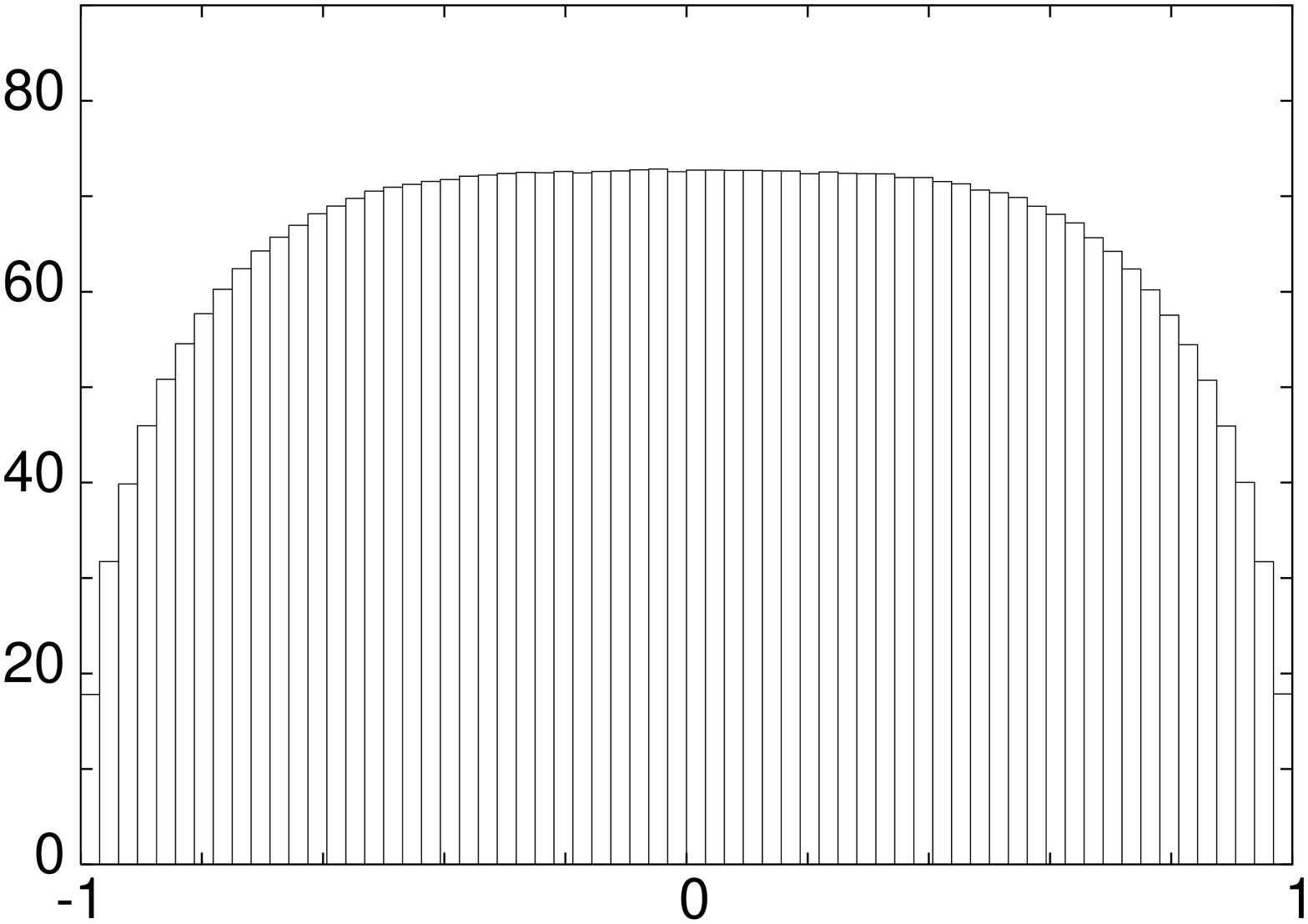}}
\subfigure[$\beta_A=1.4$]{
\includegraphics[width=2.75cm,angle=-90]{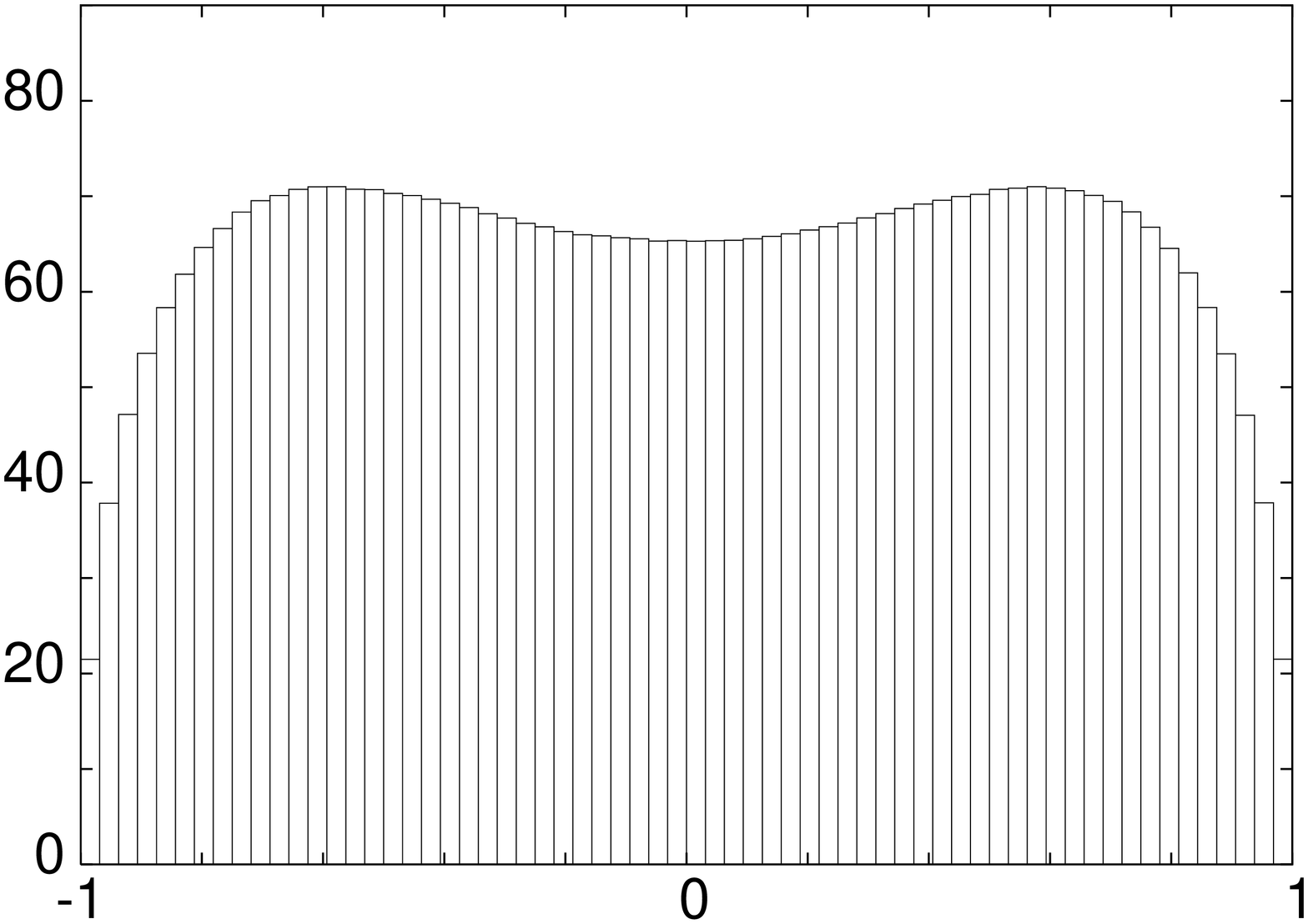}}
\caption{Check of the twist during the simulation (a); distribution of the fundamental Polyakov loop variable $L_F(\vec{x})$ in the trivial twist
sector at different values of $\beta_A$ for $V=4\times 16^3$,$\lambda=1.0$ (b-d).}
\label{fig:spadi}
\end{figure}

We focused our attention on the case $\lambda=1.0$ and we used various initial 
conditions, with trivial ($k_x=0,k_y=0,k_z=0$) and 
non-trivial twist. We monitored the twist during the runs and found
that at least for the volume we used ($V=4\times 16^3$) it did not change;
an example of such a check is given in Fig.\ref{fig:spadi} (a) for 
$\beta_A=0.9$. Similar plots hold for all the values of 
$\beta_A$ we used.
For trivial twist the distribution of the fundamental Polyakov loop
variable $L_F(\vec{x})$ is seen to
change the shape by varying $\beta_A$, 
supporting the idea of a finite temperature phase 
transition close to $\beta_A=1.2$ (see Fig.\ref{fig:spadi}).

\section{SYMMETRY AND ORDER PARA\-METER}

It is important to understand the symmetry breaking mechanism, if any,
in order to define an order parameter, allowing a quantitative study of the phase
transition and offering some insight into the mechanism of confinement.
The only hints we have are the change in the distribution of the Polyakov loop 
and the values it takes in the continuum limit.
After maximal abelian gauge (MAG) \cite{MAG} and abelian projection it is indeed possible to establish a  global symmetry which can be broken at the phase transition and a related order parameter. In the general case 
we looked for a symmetry operator $P(\vec{x})\in SO(3)$ satisfying the
following conditions:\\
(1) acting on the temporal links at a fixed time-
    \hspace*{.5cm}slice $t_1$, i.e.\\
 \hspace*{1.8cm}$\tilde{U}_{4}(\vec{x},t_1)=P(\vec{x})U_{4}(\vec{x},t_1)$ 
 \hspace{.5cm} $\forall \vec{x}$\\
    \hspace*{.45cm} and leaving the plaquette action invariant; \\
(2) mapping configurations with  $L_A\simeq 1$ into 
    \hspace*{.45cm} configurations with $L_A\simeq -\frac{1}{3}$ and vice-versa. \\
Condition (2) implies $P^2(\vec{x})=\mathbb{I}_3$; the only solutions
of this equation are:
$
\begin{array}{cccc} 
  P(\vec{x})= & \mathbb{I}_3,    & \mathbb{I}_3+2(\hat{n}(\vec{x})\cdot\vec{T})^2 & \textrm{in}\;\; SO(3),\\
  P(\vec{x})= & \pm\mathbb{I}_2, & \pm i \hat{n}(\vec{x})\cdot\vec{\sigma}        & \textrm{in}\;\; SU(2),
\end{array}
$
with $|\hat{n}|=1$, $\vec{\sigma}$ and $\vec{T}$ are the
generators of the $SU(2)$ algebra in the fundamental and in the adjoint
representation, respectively.\\
If $P(\vec{x})=\mathbb{I}_3+2(\hat{n}(\vec{x})\cdot\vec{T})^2,
\pm i \hat{n}(\vec{x})\cdot\vec{\sigma}$ it can always be decomposed
as $ P(\vec{x})=\Omega^\dagger(\vec{x})J_3\Omega(\vec{x}) $
where $\Omega(\vec{x})$ is a generic group element and
$J_3=\mathbb{I}_3+2T_3^2$ for $SO(3)$, 
$J_3=i\sigma_3$ for $SU(2)$.
The requirement of the invariance of the plaquette implies 
\begin{eqnarray}
\mathrm{Tr}[U_i(\vec{x},t_1)U_4(\vec{x}+\hat i,t_1)
     U_i^\dagger(\vec{x}+\hat4,t_2)U_4^\dagger(\vec{x},t_1)]= \nonumber\\
\mathrm{Tr}[U_i(\vec{x},t_1)P(\vec{x}+\hat i)U_4(\vec{x}+\hat i,t_1)U_i^\dagger(\vec{x}+\hat4,t_2)\cdot\nonumber\\
U_4^\dagger\!(\vec{x},t_1)\!P^\dagger\!(\vec{x})]\,, \hspace{.2cm}
\forall \vec{x},i=1,2,3,t_1\mathrm{fixed},t_2=t_1+1.\nonumber 
\end{eqnarray}
A sufficient condition which satisfies the previous equation is given by\\
$
P(\vec{x})=\frac{1}{3}\sum_{i=1}^3 [U_i(\vec{x},t_1)P(\vec{x}+\hat i)U_i^\dagger(\vec{x},t_1)+\\
U_i^\dagger(\vec{x}\!-\!\hat i,t_1)P(\vec{x}\!-\!\hat i)U_i(\vec{x}\!-\!\hat i,t_1)] \,,\hspace{.5cm}
\forall \vec{x}\;,\;t_1 \:\mathrm{fixed}.
$
It is straightforward to show that it is the global extremum of 3D MAG condition 
\cite{Poly}. 
After the implementation of a 3D MAG $P(\vec{x})$ reduces to $J_3$ and one can
transform all the t-links at a fixed time-slice as
$\widetilde{U}_4(\vec{x},t_1)\equiv J_3U_4(\vec{x},t_1)$
in order to define a modified Polyakov loop $\widetilde{L}_A$ and a modified action
$\widetilde{S}$, where the links $U_4(\vec{x},t_1)$ are substituted by
$\widetilde{U}_4(\vec{x},t_1)$.

If it would be a true symmetry it should leave the action invariant.
The numerical check shows, that the symmetry is approximately realized at the
level of $1\div 2 \%$.\\
$\langle(S\!\!-\!\!\widetilde{S})/S\rangle=0.011\;\;(\!V\!=\!4\!\!\times\!\!10^3,\beta_A\!=\!0.9,\lambda\!=\!1.0)$,\\
$\langle(S\!\!-\!\!\widetilde{S})/S\rangle=0.018\;\;(\!V\!=\!4\!\!\times\!\!10^3,\beta_A\!=\!1.6,\lambda\!=\!1.0)$.

In this way an order parameter can be defined 
\begin{equation}
\hspace*{2cm}\Delta=\frac{3}{4}|L_A-\widetilde{L}_A|
\end{equation}
interpolating between 0 ($\beta_A=0$) and 1 ($\beta_A\to\infty$).
A preliminary investigation shows that it increases by increasing $T$ and it approaches
0 faster for higher volumes at $\beta_A\lesssim 1.2$.
\begin{figure}
\includegraphics[angle=-90,width=7.5cm]{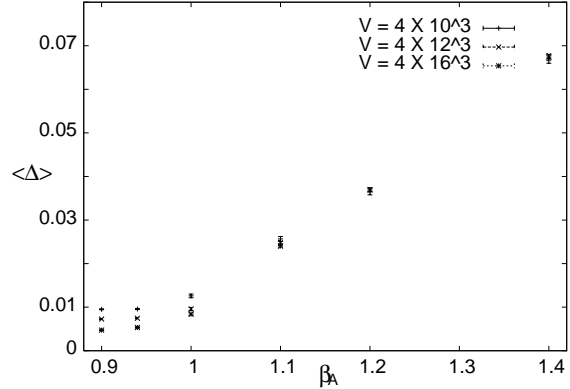}
\caption{Ensemble average of $\Delta$ vs. $\beta_A$}
\end{figure}

\section{CONCLUSIONS}

We studied the phase diagram of the mixed fundamental-adjoint action with a chemical potential
which suppresses the $Z_2$ magnetic monopoles in order to 
decouple the unphysical phase transition from the finite temperature phase transition.
A first indication of a finite temperature phase transition is given by the behaviour of the distribution of the Polyakov loop variable $L_F(\vec{x})$. We found, after 3D MAG, that $P=J_3$ generates an
approximate symmetry of the action which seems to be spontaneously broken at the 
phase transition
and can be used to define an order parameter.
This ongoing work was funded by the DFG-GK 271.

\end{document}